
\input PHYZZX
\date{September 1992}    
\Pubnum{\caps UPR--527--T}

\def\to{\rightarrow}

\titlepage
\title{Domain Walls in $N=1$ Supergravity\foot{Talk given
at the International Symposium on Black Holes, Membranes, Wormholes, and
Superstrings,
The Woodlands, Texas, January 1992.} }

\frontpageskip=0.5\medskipamount plus 0.5 fil
\author{ Mirjam Cveti\v c\foot{email CVETIC@cvetic.hep.upenn.edu}
and Stephen Griffies\foot{email GRIFFIES@cvetic.hep.upenn.edu} }
\address{Department of Physics\break
        University of Pennsylvania\break
        209 So. 33rd Street \break
        Philadelphia, PA 19104--6396}

\abstract{ We discuss a study of domain walls
in $N=1, d=4$ supergravity.  The walls saturate the Bogomol'nyi bound of
wall energy per unit area
thus proving stability of the classical solution.
They interpolate between two
vacua whose cosmological constant is non-positive and in general
different.
The matter configuration and induced
geometry are static.
We discuss the field theoretic realization of these walls
and classify three canonical configurations
with examples.
The space-time induced by a wall interpolating between the
Minkowski (topology $\Re^{4}$)
and anti-de~Sitter (topology $S^{1}(time) \times \Re^{3}(space)$)
vacua is discussed.}

\singlespace

\chap{Introduction}

Global and local topological defects are known to arise
during symmetry breaking phase transitions if
the vacuum manifold is not simply connected. Textures,
monopoles, strings, domain walls and combinations
thereof are examples.  These objects
may have important physical implications, especially in
the cosmological context.

\REF{\DGHR}{
A. Dabholkar, G.W. Gibbons, J.A. Harvey,
and F.R. Ruiz,
Nucl. Phys. \bf B340 \rm (1990) 33.}

\REF{\VAFA}{B. Greene, A. Shapere,
C. Vafa and S.-T. Yau, Nucl. Phys. \bf B337 \rm (1990) 1.}

\REF{\REY}{
S.-J. Rey, \sl Axionic String Instantons and Their
Low-Energy Implications, \rm `Particle Theory and Superstrings', ed.
L. Clavelli and B. Harm, World Scientific Pub. Co. (1989);
Phys. Rev. \bf D43 \rm (1991) 526;
\sl On String Theory and Axionic Strings
and
Instantons, \rm talk given at `Particle \& Fields `91', Vancouver Canada
(1991); \sl Exact N=4 Superconformal
Field Theory of Axionic Instantons, \rm SLAC-PUB-5662 (1991).}

\REF{\OT}{
B. Ovrut and S. Thomas ,
\sl Instantons in Antisymmetric Tensor Theories in Four-Dimensions,\rm
UPR-0465T, (March 1991),
and
Phys.Lett.{\bf B}267 (1991) 227.}

\REF\ANDY{A. Strominger,
Nucl. Phys.
\bf B343 \rm (1990) 167;
erratum \sl ibid \bf B353 \rm (1991) 565.}

\REF{\CHS}{
C. G. Callan, Jr. J. A. Harvey  and A. Strominger
Nucl. Phys. {\bf B359} (1991) 611.}

\REF{\CHSII}{
C. G. Callan, Jr. J. A. Harvey  and A. Strominger,
\sl Worldbrane Actions for String Solitons \rm , PUPT-1244(March 1991).}

\REF{\DUFFLU}{ M. J. Duff and J. X. Lu, Phys. Rev. Lett. \bf 66 \rm
(1991)1402; Nucl. Phys. \bf B354 \rm (1991) 129;
\bf B354 \rm (1991) 141; \bf B357 \rm (1991) 534; Class. Quant. Grav.
\bf 9 \rm (1992) 1; Phys. Lett. \bf B273 \rm (1991) 409.}

The inclusion of gravity in the study  of
topological defects is straightforward and
usually leads to insignificant modifications to the otherwise
stable topological defects.
However, in superstring
theories, for example, gravity and other moduli and matter fields
are on an equal footing so the effects of gravity can yield
qualitatively different features.  With the advent
of deeper understanding of semi-classical superstring theories
in a topologically nontrivial sector, various
stringy topological defects were discovered:
stringy cosmic strings \refmark{\DGHR\ , \VAFA},
axionic
instantons \refmark{\REY\ , \OT\ }
as well as
related heterotic
five-branes and other solitons\refmark{\ANDY\ , \CHS\ , \CHSII\ , \DUFFLU}.

\REF\CQR{
M. Cveti\v c, F. Quevedo and S.-J. Rey,
Phys. Rev. Lett. \bf 63 \rm (1991) 1836.}

\REF{\AT}{ E. Abraham and P. Townsend, Nucl. Phys. {\bf 351B}(1991) 313.}

\REF{\FILQI}{A. Font, L.E. Ib\'a\~nez,
D. L\"ust and  F. Quevedo, Phys. Lett. {\bf 245B}
(1990) 401;
S. Ferrara, N. Magnoli, T.R. Taylor and
G. Veneziano, Phys. Lett. {\bf 245B}
(1990) 409;
H.P. Nilles and M. Olechowski, Phys. Lett. {\bf 248B}
(1990) 268;
P. Binetruy and M.K. Gaillard, Phys. Lett. {\bf 253B}
(1991) 119.}

\REF{\CFILQ}{M. Cveti\v c, A. Font, L.E. Ib\'a\~nez,
D. L\"ust and F. Quevedo, Nucl. Phys. \bf B361 \rm (1991)
194.}

The above solutions were known to exist for free moduli fields, \ie
vanishing superpotential. Additionally, there exist
supersymmetric  domain walls when a nontrivial
superpotential for the moduli fields  exists\refmark{\CQR\ , \AT }.
These domain walls  are
interesting by themselves as well as
in connection to the dynamical supersymmetry
breaking mechanism in superstring theory\refmark{\FILQI\ , \CFILQ}.
Additionally, they
serve as a class of stringy
topological defects in which a nonzero superpotential
is essential to their existence.

The present discussion centers on the construction and properties
of domain walls in $N=1, d=4$ supergravity.
There are three major results of our analysis.
The first is a proof of a
positive energy density theorem
for a topologically nontrivial extended object
in which the matter part of the theory has a generic nonzero superpotential.
To the best of our knowledge, the proof has not been addressed previously.

\REF{\IPSK}{
A. Vilenkin, Phys. Lett. \bf 133 B \rm (1983)
177; J. Ipser and P. Sikivie, Phys. Rev. \bf D30 \rm (1984) 712.}

The second result is an existence proof for
\sl static \rm domain wall solutions for both the space-time
metric and the matter field
interpolating between two supersymmetric vacua.
It is known
that the inclusion of gravity to reflection symmetric domain walls
of infinite extent and infinitesimal thickness generically admits
only \sl time-dependent \rm solutions to Einstein's equations\refmark{\IPSK}.
We show that by
allowing for a reflection asymmetric solution  interpolating
between either a Minkowski and anti-de~Sitter space-time
or anti-de~Sitter and anti-de~Sitter space-time,
the metric
and matter field can both be
\sl time-independent.\rm

\REF{\RELA}{See, for example, C. Misner, K. Thorne and
J. Wheeler, \sl Gravitation, \rm 1973; R.M. Wald, \sl General Relativity,
\rm 1984.}

\REF{\CGRII}{ M. Cveti\v c, S. Griffies, and S.-J. Rey, \sl Nonperturbative
Stability of Supergravity and Superstring Vacua, \rm
NSF-ITP-92-84, UPR-494-T, YCTP-P44-91
(May 1992), Nucl. Phys. {bf}, in press.
. }

\REF{\GHW}{
G.W. Gibbons, C.M. Hull and  N.P. Warner,
Nucl. Phys. \bf B218 \rm (1983) 173 ;
C.M Hull, Nucl. Phys \bf B239 \rm (1984) 541.}

The last result  is that
supersymmetric domain walls
can interpolate between
two vacua of different scalar potential energy: for example,
between a supersymmetric vacuum with zero cosmological constant
(Minkowski space-time)
and another with a negative cosmological constant (anti-de~Sitter
space-time).
This result is at first counter to the notion of a domain wall interpolating
between degenerate vacua.  The point is that in defining degenerate energy
solutions,
one must include all the relevant energy in the theory; in this
case both matter and gravity.  It turns out that when the
vacua of the theory preserve supersymmetry, their energy, which is
defined in the appropriate way according to the ADM
prescription\refmark{\RELA},
are the same regardless of the particular matter vacuum energy.
This result is consistent with there being no semi-classical
tunnelling bubble causing the decay of one supersymmetric vacua
into another with a lower matter vacuum energy.
In \refmark{\CGRII} this result has been proven by showing the minimum
energy bubble which one could conceivably form separating
two supersymmetric vacua has an infinite radius and thus
will never form.  This result is complementary to positive energy
theorems
(see, for example
\refmark{\GHW} and references therein)
derived to show the stability of supergravity theories
with a matter potential unbounded below.
Indeed, without this result, one could never expect to find
the domain wall solutions we wish to describe.

\REF{\RINDLER}{W. Rindler, \sl Essential Relativity \rm , Springer-Verlag,
Berlin, 1979.}

\REF{\CGRI}{ M. Cveti\v c, S. Griffies, and S.-J. Rey, Nucl. Phys.
\bf B381 \rm (1992), 301.}

\REF{\CG}{M. Cveti\v c and S. Griffies, Phys. Lett. \bf 285B \rm (1992), 27.}

\REF{\CGS}{M. Cveti\v c, R. Davis, S. Griffies, and H. H. Soleng, work in
progress.}

This paper is organized as follows.
We start in chapter 2 with a discussion of
the formal aspects
of the realization of these
walls in the supergravity theory.
Chapter 3 gives a classification of the walls
in terms of the various combination of vacua that they
interpolate between.
Chapter 4 presents examples of the three canonical wall
configurations and chapter 5 gives the geodesic
structure for the space-time induced by these walls.
We finish with some further remarks on the wall
interpolating between Minkowski and anti-de~Sitter
space-times.

Most of the work presented here is developed
in the following references.
The field theoretic results can be found in
reference\refmark{\CGRI}.  Additional work addressing the problem
of the stability of supersymmetric vacua
can be found in
reference\refmark{\CGRII}. Discussion of the classification of the types of
walls and their  geodesic structure can be found in reference\refmark{\CG}.
Finally, the causal structure of the
Minkowski-AdS wall as well as some phenomena related to quantum fields
on this background is work in progress\refmark{\CGS}.

\REF{\WB}{J. Wess and J. Bagger, \sl Supersymmetry and Supergravity \rm ,
2nd edition, Princeton University Press, 1991.}

\chap{Supergravity realization of the walls}

Consider an $N=1$ locally supersymmetric theory with one chiral
matter superfield $\cal T$.
The bosonic part of the $N=1$ supergravity Lagrangian
is\refmark{\WB}\foot{We do not choose
the commonly used  K\"ahler gauge
which introduces the potential function\refmark{\WB}
$G(T,\bar T) = K(T,\bar T) + ln|W(T)|^{2}$, since it is not
adequate for situations in which the superpotential is allowed to vanish.}
$$
e^{-1}L = -{1 \over 2 \kappa}R + K_{T \bar T}g^{\mu \nu}\partial_{\mu}\bar T
\partial_{\nu}T - e^{\kappa K}(K^{T \bar T}|D_{T}W|^{2} - 3 \kappa |W|^{2})
\eqn\localL$$
where
$e = |detg_{\mu \nu}|^{1 \over 2},
K(T, \bar T) =$ K\"ahler potential and $D_{T}W \equiv e^{-\kappa K} (\partial_T
e^{\kappa K} W)$.
\foot{We use
the conventions: $\gamma^{\mu}=e^{\mu}_{a}\gamma^{a}$ where
$\gamma^{a}$ are the flat spacetime Dirac matrices satisfying
$\{\gamma^{a},\gamma^{b}\}=2\eta^{ab}$; $e^{a}_{\mu}e^{\mu}_{b}
= \delta^{a}_{b}$; $a=0,...3$; $\mu=t,x,y,z$; $\overline{\psi} =
\psi^{\dagger}\gamma^{t}$; $(+,-,-,-)$ space-time signature;
and write $\kappa =
8\pi G_{N}$.}

In order to have stable domain wall solutions,
topological arguments imply that the degenerate vacua
be disconnected.
Thus
one must have isolated
vacua of the matter potential. However, the inclusion of gravity
will turn out to play an important role in removing the
constraint  that the isolated minima of the \sl matter \rm potential have to
be degenerate.  We shall see that with the inclusion of gravitational energy,
the notion
of degenerate vacua will be defined as
\sl supersymmetry preserving vacua \rm
just as in globally supersymmetric theories.
Indeed, formal arguments for the stability follow from the existence of
local supersymmetry charges which satisfy an
algebra which is a generalization of the global algebra.
Thus, the inclusion of gravity, when the dust settles, merely adds to the
technology
necessary to formulate the existence and stability criteria of these extended
objects.
Therefore, in a formal sense, the arguments are analogous to those in the
global case\refmark{\CQR\ , \AT\ , \CGRI}.

Supersymmetry preserving minimum of the  potential
in \localL\ satisfy
$D_{T}W = 0$. This in turn implies
that the supersymmetry preserving vacua have
either  zero cosmological constant (Minkowski space-time)
when $ W=0$, or negative  cosmological constant
$-3e^{\kappa K}|\kappa W|^2$  (anti-de~Sitter space-time) when
$W \not=0$.
Note that we define the cosmological
constant as follows.
The energy momentum tensor when $T$ is at its vacuum value
($D_{T}W = 0)$ is
$T_{\mu \nu} = -3\kappa|We^{{\kappa K \over 2}}|^{2}g_{\mu \nu}$.
Therefore,
Einstein's equation $R_{\mu \nu} - {1 \over 2}g_{\mu \nu}R
= \kappa T_{\mu \nu}$ can be written
$R_{\mu \nu} - {1 \over 2}g_{\mu \nu}R  = \Lambda g_{\mu \nu}$
with $\Lambda = -3|\kappa W e^{{\kappa K \over 2}}|^{2}$.

\REF{\WITTEN}{E. Witten, Comm. Math. Physics,
\bf 80 \rm (1981) 381.}

\REF{\NESTER}{J. M. Nester, Phys. Lett. \bf 83A \rm (1981) 241.}

\REF\DES{S. Deser, Class. Quantum. Grav. \bf 2 \rm
(1985) 489.}

\section{ADM Mass Density}

In the following
we obtain a lower bound on the
mass density of domain walls
living in this theory.
In that regard, we employ the results
of  \refmark{\WITTEN}
and  \refmark{\NESTER}
who addressed the positivity of the ADM
mass in general relativity, as well as certain
generalizations to  anti-de~Sitter backgrounds\refmark{\GHW}. We note that
the ADM mass
for spatially infinite objects is
not well-defined\refmark{\DES}.
However, as a weaker requirement, we will assume that
the ADM procedure is valid for the mass per unit area rather than the
mass of the domain wall. Indeed, this is the energy which is of interest
since the total mass is, by definition, infinite.

Consider the supersymmetry charge density
$$
Q[\epsilon'] = \int_{\partial \Sigma} \bar{\epsilon'}
\gamma^{\mu \nu \rho} \psi_{\rho} d\Sigma_{\mu \nu}
\eqn\localcharge$$
where $\epsilon'$ is a commuting Majorana spinor,
$\psi_{\rho}$ is the spin $3/2$ gravitino field, and
$\Sigma$ is a spacelike hypersurface.
Taking a supersymmetry variation of $Q[\epsilon']$ with respect to
another commuting Majorana spinor $\epsilon'$ yields
$$
\eqalign
{
\delta_{\epsilon} Q[\epsilon']& \equiv \{Q[\epsilon'],
\bar{Q}[\epsilon]\}  \cr
&= \int_{\partial \Sigma}N^{\mu \nu} d\Sigma_{\mu \nu}
= 2\int_{\Sigma}\nabla_{\nu}N^{\mu \nu} d\Sigma_{\mu}  \cr
}
\eqn\localchargevariation$$
where
$N^{\mu \nu} = \bar \epsilon'\gamma^{\mu \nu \rho}
\hat\nabla_{\rho} \epsilon $ is a generalized
Nester's form\refmark{\NESTER}.  Here
$\hat\nabla_{\rho}\epsilon \equiv
\delta_{\epsilon}\psi_{\rho} =
[2\nabla_{\rho} + ie^{K \over 2}(WP_{R} + \bar{W}P_{L})\gamma_{\rho}
 - Im(K_{T}\partial_{\rho}T)\gamma^{5}]\epsilon$ and
 $\nabla_{\mu}\epsilon = (\partial_{\mu}
  + {1\over2}\omega^{ab}_{\mu}\sigma_{ab})\epsilon$.
 In \localchargevariation\
the last equality follows
from Stoke's law.

We consider an Ansatz for
the space-time metric
$ds^{2} = A(z,t)(dt^{2} - dz^{2}) +\break
 B(z,t)(-dx^{2} - dy^{2})$
characteristic of space-times with a domain wall where $z$ is
the coordinate transverse to the wall.
However, we do not assume \sl a priori \rm
that the metric is symmetric about the plane $z=0$. Nor do we
assume a particular behavior of $A$ and $B$ at
$|z|\rightarrow\infty$ except that the asymptotic metric satisfies
the vacuum Einstein equations  with a zero or negative cosmological
constant.

We are concerned with supercharge \sl density \rm and thus
insist upon only $SO(1,1)$ covariance in the $z$ and $t$ directions.
This in turn implies
that the space-like
 hypersurface
$\Sigma$ in eq.\localchargevariation\
is the $z-$axis
with measure
$d\Sigma_{\mu} = (d\Sigma_{t},0,0,0) =
|g_{tt}g_{zz}|^{1\over 2}dz$.  The boundary
$\partial \Sigma$ are then the
two asymptotic points $z\rightarrow \pm\infty$.
Technical details in obtaining the explicit form of
eq.\localchargevariation\
are given in reference \refmark{\CGRI}
and will be omitted here.

After some algebra, the volume integral
yields:
$$
2\int_{\Sigma}\nabla_{\nu}N^{\mu \nu} d\Sigma_{\mu}=
\int_{-\infty}^{\infty}
[-\delta_{\epsilon'}\psi^\dagger_i g^{ij} \delta_{\epsilon}\psi_j +
K_{T \bar T}\delta_{\epsilon'}\chi^\dagger \delta_{\epsilon}\chi]dz
\eqn\volumeintegral$$
where $\delta_{\epsilon}\psi_{i}$ and $\delta_{\epsilon}\chi$
are the supersymmetry variations of the fermionic fields in the
bosonic backgrounds.
Upon setting $\epsilon' = \epsilon$ the expression \volumeintegral\
is a positive definite quantity which in turn (through
 eq.\localchargevariation\ )
yields  the bound
$\delta_{\epsilon} Q[\epsilon] \ge 0.$

Analysis of the surface
integral in \localchargevariation\
yields two terms: $(1)$ The ADM mass density of
the configuration,
denoted $\sigma$ and
$(2)$ The topological charge density,
denoted $C$.
Positivity of the volume integral
translates into the bound
$$
\sigma \ge
 |C|
\eqn\localbound$$
which is saturated iff $\delta_\epsilon Q[\epsilon]=0$. In this
case the bosonic backgrounds are supersymmetric; \ie\
they satisfy
$\delta\psi_{\mu} = 0$   and
$\delta\chi = 0$ (see eq.\volumeintegral\ ).
Such configurations saturate the previous bound thus establishing their
stability.

\REF{\BOGO}{E. B. Bogomol'nyi, Sov. J. Nucl. Phys. \bf 24 \rm (1976), 449.}

\section{Self-Dual Equations}

We now concentrate on solving for the space-time metric
and matter field configuration in the supersymmetric
case.  This calculation involves an analysis of
the first order equations
$\delta_{\epsilon} \psi_\mu = 0$ and $\delta_{\epsilon}
\chi = 0$ which is discussed in \refmark{\CGRI}.
\foot{We call these first order differential equations
the Bogomol'nyi\refmark{\BOGO} or self-dual equations.
Their square gives
the classical equations of motion.}
The self-dual equation
for the matter field $T(z)$ follows from
$\delta_{\epsilon} \chi = 0$:
$$
\partial_z T(z) = ie^{i \theta}\sqrt{A}
e^{K\over 2}K^{T \bar T}D_{\overline{T}}\overline{W}
\eqn\localbogeq
$$
with a constraint on the $\epsilon$-spinor:
$$
  \epsilon_1=e^{i\theta}\epsilon_2^*.
\eqn\localspineq
$$
The undetermined phase
$e^{i\theta}$ is in general a space-time
coordinate-dependent function.

Since we wish to define the ADM mass per unit area
of the domain
wall unambiguously, we
look for a \sl time-independent \rm metric solution.
For the walls studied
in \refmark{\IPSK}, the resulting reflection symmetric metric is
time-dependent
even though
the energy-momentum tensor of the domain wall is time-independent
(unless one takes a special value of mass to tension ratio that is not
realized by generic field theory examples).
With no assumed reflection symmetry of the space-time metric,
\sl a priori \rm
one cannot say if there exist nontrivial time-independent
domain wall solutions.
However,
in order for our assumption of the
time
independence of the $T$-field to be consistent with the Bogomol'nyi
equation \localbogeq\ ,
the metric component $A$ \sl must \rm be time-independent.

The self-dual equations for the metric components, following from
$\delta_{\epsilon}\psi_t
= \delta_{\epsilon} \psi_x = 0$,
are
$$
\partial_{z}A^{-1/2} =
\partial_{z}B^{-1/2} =
-\kappa (ie^{-i\theta}) W e^{\kappa K \over 2}.
\eqn\metriceq
$$
Since the metric functions
$A$ and $B$ are real, the phase $e^{i\theta}$ is required to meet
a local constraint
$$
W = -i\zeta e^{i\theta}|W|
\eqn\thetaconstraint$$ where $\zeta = \pm$.
Assuming continuity, $\zeta = \pm$ can
change
only at points where
$W$ vanishes.
This connection between the metric and matter superpotential
restricts the possible $W$ admitting walls in the local theory.
This result is in contrast to the global case in which all $W$ with
degenerate vacua admit wall solutions.
We comment on this result later.

$\delta_{\epsilon} \psi_z=0$ yields the differential equation
for the $z$ dependent phase $\theta$:
$$
\partial_{z} \theta = - Im(K_{T}\partial_{z}T).
\eqn\otherthetaconstraint$$
Consistency of \localbogeq\ , \metriceq\ and \otherthetaconstraint\
with
\thetaconstraint\ leads to the following sufficient conditions for the
existence of a static supersymmetric domain wall:
$$
Im(\partial_{z}T{D_{T}W\over W}) = 0,
\eqn\summaryI$$
$$
\partial_z T(z) = -\zeta \sqrt{A}|W|
e^{\kappa K \over 2}K^{T \bar T}
{D_{\overline{T}}\overline{W}\over \overline{W}},
\eqn\summaryII$$
$$
\partial_{z}A^{-1/2} =
\partial_{z}B^{-1/2} =
\kappa \zeta \sqrt{A}|W|e^{\kappa K \over 2},
\eqn\summaryIII$$
as well as the
explicit expression for the ADM mass density (energy per
area or surface tension)
of the supersymmetric domain wall
configuration
$$
\sigma = |C| \equiv 2 |(\zeta|We^{\kappa K \over 2}|)_{z=+\infty}
-(\zeta|We^{\kappa K \over 2}|)_{z=-\infty}|
\equiv {2 \over \sqrt{3}}\kappa^{-1}|\Delta(\zeta|\Lambda|^{1/2})|
\eqn\admenergy$$
where
 $\Lambda\equiv -3|\kappa W e^{{\kappa K \over 2}}|^{2}$
is the cosmological constant for the supersymmetric
vacuum.

We now comment on these equations.

It follows from
\admenergy\ that there are
no static domain walls saturating the Bogomol'nyi bound
that interpolate  between
two supersymmetric vacua with
zero cosmological constant.
In this case
$W(+\infty)=W(-\infty)=0$
and thus there is no
energy associated
with such a domain wall since $|C|\equiv 0$.
This result is in agreement with the results
of reference \refmark{\IPSK}, where for infinitesimally
thin domain walls with
asymptotically Minkowski space-times only
time-dependent  metric solutions were obtained.
The result from \admenergy\ implies that static
supersymmetric domain wall  solutions  exist only
if at least one of the vacua is AdS.

\topinsert
Figure 1:  The path in superpotial space traversed as the
scalar field interpolates between degenerate vacua.
The wall is realized in both the global and local theories
for path (A) and just for the global theory in path (B).

\line{\hrulefill}
\endinsert
Eq. \summaryI\  is a consistency constraint which
specifies the \sl geodesic path \rm
between two supersymmetric vacua
in the
supergravity potential space $e^{{\kappa K \over 2}} W \in \bf C$ when
mapped from the
$z$-axis $(-\infty, + \infty)$.
This geodesic equation
has qualitatively new features
in comparison with
the geodesic equation in the
global supersymmetric case\refmark{ \CQR\ , \AT\ , \CGRI}.
While in the global case
geodesics
are arbitrary straight lines in the $W-$plane,
the local
geodesic equation in
the limit $\kappa \to 0$ (global
limit of the local supersymmetric theory)
leads to the geodesic equation
$Im({\partial_{z}W\over W})\equiv
\partial_z \vartheta = 0$
where $W$ has been written
as $W(z) =|W|e^{i\vartheta}$.
This in turn implies that
as $\kappa \rightarrow 0$ the local geodesic equation
reduces to the constraint that
$W$
has to be  a  straight line passing through the
origin;
\ie\ the phase of $W$ has to be
constant mod $\pi$. Figure 1 illustrates these points.
This observation in turn implies that
the introduction of gravity
imposes a strong constraint on the type of
domain wall solutions realized. In particular,
domain wall solutions in the global case interpolating
between vacua
in the $e^{{ \kappa K \over2}}W$ plane
that do not
lie along a straight line passing through the origin
\sl do not \rm have an analogous solution in the local case.
This result is a
manifestation of the singular nature of a perturbation
in Newton's constant as seen in \summaryIII\ .
Another way to understand the inability of all global
walls to be realized in the local theory is that the
space-time metric introduces an extra field degree of
freedom to the local
theory which allows for an extra direction to connect previously
disconnected vacua.

Eq. \summaryII\ for the $T$ field
(the ``square root'' of the equation of motion for $T$)
and eq. \summaryIII\
for the metric (the ``square root'' of the Einstein's equation)
are invariant under $z$ translation as well as
under rescalings of $(A,B)\to \lambda^2 (A,B)$
and $z\to\lambda^{-1}z$.
Additionally,
eq. \summaryII\
implies that $\partial_z T(z)
\rightarrow 0$ as one approaches the supersymmetric minima which are
points where $D_T W=0$, thus indicating
a solution smoothly interpolating between supersymmtric vacua.
In general, the field $T$ reaches the supersymmetric minimum
exponentially fast as a function of $z$.

\chap{Classification of the Walls}

We now concentrate the equation \summaryIII\
for the metric.
Our aim is
to classify all  the
qualitatively different metric
configurations.
First, we
set $A(z)=B(z)$ without loss of generality
which implies that the metric is
conformally flat.
Also, we emphasize in
\summaryIII\
the singular limit
when gravity is turned off ($\kappa \rightarrow 0$).
As noted earlier, the same  singular limit ($\kappa\to 0$)
is also responsible for the  restrictive
geodesics in the $W$-plane  compared to
a global theory
which  contains no gravitational information
($\kappa =0$).
For $\kappa =0$, the conformal factor
factor $A$ is  constant in the whole space; \ie\ we have
flat space-time everywhere.  However, the
moment $\kappa > 0$, $A$ varies  with $z$.
Thus, our aim is to study the
nature of the conformal factor $A(z)$.
We classify
three types of static domain wall configurations
which depend on the
nature of the potential of the matter field.
For illustrative purposes to indicate the nature of the
minima between which a wall interpolates, we sketch a
typical scalar potential in Figure 2.  The non-degeneracy, as
emphasized throughout the paper, is deceptive since degeneracy
is based on both gravity and matter energy; the scalar potential
only involves the matter part.

\topinsert
Figure 2: The projection of a
typical scalar potential on a
particular complex $T$ direction
indicating the supersymmetric
minima ($D_{W}T = 0$) between which the domain wall
interpolates.  These minima are in general not degenerate.
However, since they are supersymmetric, when gravitational
energy is included they become energetically the same thus allowing for
domain wall solutions.

\line{\hrulefill}
\endinsert

$(I)$ A wall interpolating between
a supersymmetric AdS
vacuum ($|W_{+\infty}|\ne 0$)
and a
Minkowski supersymmetric
vacuum ($|W_{-\infty}|=0$).
 From \summaryIII\ one sees that  on the Minkowski
side
the conformal factor approaches a
constant which
can be normalized to unity; \ie\
$$A(z)\to 1,\ \   z\to -\infty. \eqn\minm
$$
On the
AdS side $A(z)$ falls off as
$z^{-2}$  with the
strength of the fall-off  determined by the strength of
the cosmological constant; \ie\
$$A(z)\to
{ 3\over
{|\Lambda_{+\infty}| z^2}},\ \
z \to +\infty. \eqn\adsm
$$
The surface energy of this configuration as determined from \admenergy\
is
$$\sigma_{I} = {2 \over \sqrt{3}}\kappa^{-1}|\Lambda_{+\infty}|^{1/2}.
\eqn\sigmaI
$$
Here, the cosmological constant of the supersymmetric
AdS vacuum is $\Lambda_{+\infty} =\break
-3|\kappa
W e^{{\kappa K \over 2}}|^{2}_{+\infty}$.

$(II)$ A wall interpolating between
two supersymmetric AdS vacua and where the superpotential
passes through
zero in between.
The
cosmological constant
need not be the same in both vacua.
The point where
$W=0$
can be chosen at  $z=0$ without loss of generality due to the translational
symmetry of the system.
At this  point $\zeta$ changes sign and thus
$\zeta_{+\infty}=-
\zeta_{-\infty}=1$.
 The conformal factor
has the same asympotic behaviour  on both sides of the
domain wall:
$$
A(z) \to
{ 3\over
{|\Lambda_{\pm
\infty}| z^2}},\ \
z \to  \pm \infty\eqn\madsII
$$
while at $z=0$, \ie\ when $W=0$, the conformal factor
levels out, \ie\
$\partial_z A(z)_{z=0}=0$.
In other words   $A(z)$
 has a characteristic  (in general asymmetric)
bell-like
shape.

The surface energy of this configuration is
$$\sigma_{II} = {2 \over \sqrt{3}}\kappa^{-1}
(|\Lambda|^{1/2}_{-\infty} + |\Lambda|^{1/2}_{+\infty})\eqn\sigmaII.$$

$(III)$ A wall interpolating between
two AdS vacua, while the superpotential
does \sl not \rm pass through zero. Again, the
cosmological constant
need not be the same in both vacua.
In this case, since $|W|$ is never zero,
$\zeta$ has
the same sign in the whole region, say, +1. Eq.\summaryIII\
in turn implies that
the conformal factor necessarily
blows up at some coordinate
$z^{*}$.  In general,
the matter field $T$ has long since interpolated
between the two vacua by the time the metric
reaches $z^{*}$.
Thus, the domain wall, defined as the region over which
$T$ moves from one vacuum to another, lies entirely within
the coordinate
region $
z^{*}<z<+\infty$.
The conformal factor has the asymptotic behaviour:
$$\eqalign{
  A(z)&\to
{ 3\over
{|\Lambda_{+\infty}| z^2}},\ \
z \to +\infty \cr
A(z) &\rightarrow {3\over{|\Lambda_{z^*}|(z - z^{*})^2}}
\ \ z\to z^{*}.}
\eqn\singularity$$
The surface energy of this configuration is
$$\sigma_{III} = {2 \over \sqrt{3}}\kappa^{-1}
| |\Lambda|^{1/2}_{z^{*}}
- |\Lambda|^{1/2}_{ + \infty } |.
\eqn\sigmaIII
$$
Note that
the point $z^{*}$ is an infinite proper
spatial distance away from any
other point $z > z^{*}$ since $\int dz A^{1/2} \rightarrow
$ln$|z - z^{*}|$.

In order to understand this singularity as well as the
distinctive $z^{-2}$ behaviour of the conformal factor on the
AdS side of a wall, it is appropriate at this point to
study AdS space-time in a coordinate
system which singles out the $z$ direction.
For this purpose, we consider the metric
$$
ds^{2} = (\alpha z)^{-2}(dt^{2} - dx^{2} - dy^{2} - dz^{2})
\eqn\AdSmetric$$
with $z>0$.  As noted above, this is the form of the metric
on the AdS side of the domain wall when the $T$ field has
reached its supersymmetric vacuum.  In this context, $\alpha$
is related to the cosmological constant by $\Lambda = -3\alpha^{2}$.

\REF{\BD}{ N. D. Birrell and P. C. W. Davies, Quantum Fields in
Curved Space, Cambridge 1982.}

\REF\BF{P. Breitenlohner and D. Z. Freedman, Ann. of Phys.
\bf 144 \rm (1982) 249-281.}

Eq. \AdSmetric\ is the form for the metric
describing AdS space-time  where the translational
invariance is broken in the $z$ direction.
The curvature tensor, by definition of a maximally
symmetric space-time,
satisfies
$R_{\mu\nu\sigma\rho}=\alpha^{2}(g_{\mu\sigma}g_{\nu\rho}
-g_{\mu\rho}g_{\nu\sigma})$.
One can represent
four dimensional AdS space-time
as the hyperboloid
$\eta_{A B}Y^{A}Y^{B} = \alpha^{-2}$ embedded in the five
dimensional space with flat metric $\eta^{A B}=diag(+---+)$.
We found that the following choice of coordinates
$$
\eqalign{
Y^{0} &= te^{\alpha \tilde{ z}}, \   \
Y^{1} = xe^{\alpha \tilde{z}}, \   \
Y^{2} = ye^{\alpha \tilde{z}} \cr
Y^{3} &= (\alpha)^{-1}\sinh(\alpha \tilde{z})
      - {1 \over 2}\alpha e^{\alpha \tilde{z}}(x^{2} + y^{2} -t^{2})  \cr
Y^{4} &= (\alpha)^{-1}\cosh(\alpha \tilde{z})
      + {1 \over 2}\alpha e^{\alpha \tilde{z}}(x^{2} + y^{2} -t^{2})
}\eqn\intrinsic$$
yield the metric intrinsic to the surface
$$
ds^{2} = e^{2 \alpha \tilde{z}}(dt^{2} - dx^{2} - dy^{2}) -
d\tilde{z}^{2}.
\eqn\premetric$$
This choice of intrinsic coordinates is motivated from the cosmological
form for
the metric in \sl de~Sitter \rm space (see, for example
\refmark\BD).
By choosing $z = \alpha^{-1}e^{- \alpha \tilde{z} }$
we recover the
form of the metric in \AdSmetric\ .

These coordinates
cover one-half of the AdS manifold
since $Y^{3}+Y^{4} > 0$.  By choosing ($Y^{3}, Y^{4}, z) \rightarrow
(-Y^{3}, -Y^{4}, -z)$, we cover the $Y^{3}+Y^{4}<0$ region and have the
metric \AdSmetric\ for $z<0$.
This choice
should be contrasted
with the standard
set of coordinates  respecting spherical symmetry
about an origin which
completely covers AdS space-time\refmark{\BF}.
In this case the metric has the form
$$
ds^{2} = (\alpha \cos\rho)^{-2}(dt_{c}^{2}
         - d\rho^{2} - \sin^{2}\rho (d\theta^{2}
         + \sin^{2}\theta d\phi^{2}) )
\eqn\AdSspherical$$
with $0 \le \rho <  \pi/2,\   \  0 \le \theta \le \pi,\   \
0 \le \phi <  2\pi$, and $-\pi \le t_{c} \le \pi$.

The time-like coordinate $t_{c}$ in \AdSspherical\  is
a periodic coordinate.
However, the
coordinates \intrinsic  , in which time
ranges over $-\infty < t < \infty$,  exhibit \sl no \rm
periodic structure.
What we have effectively done in choosing the planar
coordinates \intrinsic\ is to sacrifice a complete covering of
AdS for a  non-periodic time-like
variable.  The coordinates \AdSmetric\ are extendible whereas
those of \AdSspherical\ are not.

The previous discussion of the metric \AdSmetric\
now allows for a straightforward interpretation of
the singular wall (type $III$) configuration.  What we have is a domain wall
separating two distinct regions of a
generalized AdS space-time possessing a $z$ dependent
cosmological \sl parameter \rm  which \sl never \rm passes
through zero.  The singular point $z^{*}$ corresponds to the
origin $z=0$ in the metric \AdSmetric\ .  On the ``other side''
of $z^{*}$ lives an AdS space-time symmetric to the
$z>z^{*}$ side. Together these two sides completely cover
the whole of the generalzed AdS space-time just as the regions
$z>0$ and $z<0$ in the planar coordinates leading to
\AdSmetric\ cover all of AdS.

\chap{Examples}

The above  discussion of the three types of domain wallsfoot{
In Ref.~refmark{   } the stringy examples  based on the $SL(2,{bf
Z})$ duality symmetry of the string theory is also discussed.}
is illustrated by
a simple polynomial form for the superpotential,
a flat K\"ahler manifold: $K = T\bar{T}$,
and a real $T$.
We choose the superpotential
$$
W =
\gamma T[{1\over 5}T^{4} - {1\over 3}T^{2}(a^{2} + b^{2}) + a^{2}b^{2}].
\eqn\superpotential$$
where $\gamma$ is a mass dimension $-2$ parameter which we set to unity
and $a^2$ and $b^2$ are positive  dimension $2$ parameters.
Depending on the value of the parameters $a$ and $b$, the
superpotential \superpotential\
provides us with a set of theories which
accommodate  the above three classes of the domain walls.

Note that the geodesic  constraint
$Im(\partial_{z}T{D_{T}W\over W}) = 0$ is always satisfied
for $T=\bar{T}$.
The supersymmetric vacuum satisfies
$D_{T}W \equiv W_T+\kappa K_T W =0$, where
$W_{T} = (T^{2} - a^{2})(T^{2} - b^{2})$.
Thus, for $a,b<<1/\sqrt\kappa$,
the supersymmetric vacua
take place
for real values of $T$
near $\pm a, \pm b$.
Figures $3,4$ and $5$ display the conformal factor $A$ for the
these three  classes of the domain walls.
Each example corresponds to a
different choice of the parameters $a$ and $b$, which we
took for
simplicity  to be in the range
$<<1/\sqrt\kappa$.

\topinsert
Figure 3:
Type $(I)$
conformal factor $A(z)$ for a space-time with $\Lambda_{-\infty}
= 0$
(Minkowski: $z<0$) separated by a domain wall from a space-time
with $\Lambda_{+\infty}
 < 0$ (AdS: $z>0$).  The wall, \ie\
 the
region over which the matter field $T$ changes is centered
at $z=0$  and has thickness
$\approx 200$ in $ \sqrt\kappa$ units.
The superpotential \superpotential\ has parameters $a^{2}=0,
b^{2}=0.1$ and $T$
interpolates between  $T_{-\infty}= 0 = a$
and $T_{\infty}= .318 \approx b$.

Figure 4:  Type $(II)$
Conformal factor $A(z)$ for a space-time with negative
cosmological constant separated by a domain wall from its mirror image
(\ie\ a $Z_{2}$ configuration).  The wall is centered at $z=0$ and has
thickness $\approx 200$ in $\sqrt\kappa$ units.
The superpotential
\superpotential\  has parameters $a^{2}=.025, b^{2}=0.1$.
and $T$
interpolates between  $T_{\mp\infty}= \pm .1598 \approx \pm a$.

Figure 5: Type $(III)$
conformal factor $A(z)$ for a space with negative
cosmological constant
separated by a domain wall from a space with a different
negative cosmological constant.  The superpotential $W$ never passes
through a zero as $T$ interpolates from one vacuum to another.
The domain wall is centered at $z=0$ and
has thickness $\approx 200$
where   $z$ is measured in $\sqrt\kappa$ units.
The singularity is at $z^{*}
\approx -5600$.
The superpotential
\superpotential\ has parameters $a^{2}=.025, b^{2}=0.1$
and $T$
interpolates between  $T_{-\infty}= .315 \approx b $
and $T_{\infty}= .160 \approx a$.

\line{\hrulefill}
\endinsert

\chap{Geodesic Structure}

We now turn to the
study of the
geodesic structure for  the
induced space-time.  To do so, we analyze the motion of
test particles in the background of a supersymmetric domain
wall.

The motion of massless particles
is trivial since the metric is conformally flat; they
simply define the usual $45^{\circ}$ null rays in a
space-time diagram.  Particles moving in constant $z$
planes will feel no force since the conformal factor is
only a function of the transverse coordinate $z$.
In other words, the metric is invariant under $x,y$
boosts and thus without loss we can move to an inertial frame
in which there is no motion in these directions.
Therefore,
the only interesting geodesics will come from the $1+1$
metric
$
ds^{2} = A(z)(dt^{2} - dz^{2}).
$
For massive
particles, which live on time-like geodesics, we can
parametrize the motion with the
proper-time element $ds^{2}\equiv d\tau^{2} > 0$.
Rearranging the metric and introducing the conserved energy per
mass $\epsilon\equiv
A  {dt \over d\tau}$
of the particle yields the equation for the world-line
$$
({dz \over dt})^{2} + {A \over \epsilon^{2}} = 1.
\eqn\conservation$$
On a time-like geodesic,
$0 \le ({dz / dt})^{2} < 1$, and so
the turning point, \ie\
 $v \equiv dz/dt = 0$, of the motion
is where
${A / \epsilon^{2}} = 1$.

A convenient way to understand massive particle motion is to
consider a particle with a given initial coordinate velocity
$v_{o}$ at some coordinate $z_{o}$;
from \conservation\
$\epsilon$ for such a particle is
$\epsilon^{2} = A(z_{o})(1-v^{2}_{o})^{-1}$.  Equation \conservation\
can be thought of as the conservation of energy with an
effective potential $V(z)\equiv (1-v^2)
 = {A(z) \over A(z_{o})}(1-v_{o}^{2})$.
Again, points where $V(z) = 1$ are turning
points.

For particles incident upon the type I wall from the Minkowski
side, passage through to the AdS side is always allowed.
However, the reverse motion requires the initial velocity
to satisfy $v^{2} > 1 - A(z_{o})$; otherwise there is a turning point
and the particle returns to the AdS side.  Motion in the other
domain wall backgrounds is analogous: a sketch of the
effective potential $V(z) = A(z)(1-v_{o}^{2})/A(z_{o})$
makes the motion clear.

One can understand the repulsive nature of these space-times
on the AdS side
by calculating the force on a test particle which
has a fixed position $z$ (also known as a fiducial observer).
This force can be obtained through the
geodesic equation
$ p^{\alpha}p^{\beta}_{; \alpha} = mf^{\beta} $
with $p^{\alpha} = m{dx^{\alpha} \over d\tau}$.
The gravitational force acting on the fiducial observer is
$$ f^{\beta} = ( 0, 0, 0,  -{m \over 2} A^{-2} \partial_{z}A).
\eqn\force$$
For a metric which falls off as on the AdS side of a wall, this
force is  directed
\sl towards \rm the AdS vacuum (e.g. $z=+\infty$ in the type $I$
wall depicted in figure $3$).
The magnitude of the
acceleration
is given by
$$ |a|^{2} \equiv |f_{\alpha}f^{\alpha}|/m^{2}
= ({1 \over 2}{ \partial_{z}lnA \over A^{1/2} })^{2} =
(\kappa |W|e^{{\kappa K\over 2}})^{2}.
\eqn\acceleration$$
For fiducial observers in the region where $T$ is
essentially at its vacuum value; \ie\ far away from the wall,
the proper acceleration has the
constant magnitude
$|a|^2  = |{\Lambda_{\pm\infty}/3}| $.
In this region, integration of \conservation\ yields the
hyperbolic world line for freely falling test particles
$z^{2}-t^{2}=|{\Lambda_{\pm \infty}/
 3}|^{-1}\epsilon^{-2}$.  Therefore, a
fiducial observer situated far away from a type $I$ or $II$ wall
in a
$\Lambda_{\pm\infty} \ne 0$ region will feel a constant acceleration
$|{\Lambda_{\pm\infty}/
 3}|^{1/2}$ directed away from the wall as well as
see freely falling test particles moving
away from
the wall with the a hyperbolic world line.
Such a world line is also exhibited by a particle with a
constant proper acceleration moving in Minkowski space-time.
These particles,
known as Rindler particles\refmark{\RINDLER}, are not freely
falling in the Minkowski background;
their acceleration is provided by an external
non-gravitational force.
However, we see that
AdS provides precisely this force due to the non-trivial
curvature of the space-time.
We add that on the Minkowski side of the walls, free test particles
experience no gravitational force even though there is an infinite
object nearby.

\REF{\ISR}{W. Israel, Nuovo Cimento \bf 44B \rm  (1966), 1.}

\REF{\HARALD}{{\O}. Gr{\o}n and H. H. Soleng, Phys. Lett. \bf 165A \rm ,
(1992), 191.}

One can understand the no-force result for the particles living
on the Minkowski side of the walls through the formalism of
singular hypersurfaces\refmark{\ISR}.  A straightforward calculation\foot{See
\refmark{\HARALD} for a nice example of this formalism applied to a
planar geometry.  In addition, \refmark{\IPSK} employed these
ideas in solving for the space-time around their domain walls.}
yields a
\sl negative \rm effective gravitational mass/area  due to the wall
whereas AdS has exactly the opposite positive gravitational mass.
Thus the observer on the Minkowski side of the wall   does not
feel any gravitational force.

\REF\IPSER{ J. Ipser, Phys Rev. \bf D30 \rm (1984) 2452.}

This above result should be contrasted with the
observation in
Ref.\refmark{\IPSK}, where  infinitesimally
thin reflection symmetric
domain walls with
asymptotically Minkowski space-times
always repell
the fiducial observer
with a constant acceleration $\kappa\sigma/4$.
Here, $\sigma$ is the energy per unit area of the
domain wall.
\foot{
Domain walls which separate two Minkowski vacua yet satisfy the
nonstandard relation  $\sigma = 2\tau$, where $\tau$ is the
surface tension of the wall, produce \sl no \rm gravitational force on
test particles.
Walls of isotropically and uniformly distributed
cosmic strings produce such an equation of state\refmark\IPSER .}
Recall that these domain walls always  produce
a time dependent metric.
In our case everything is static.
In particular, for the type $(I)$ domain walls
interpolating between AdS and Minkowski  space-times,
the asymptotic acceleration
on the AdS side can be written as
$a=\kappa\sigma_I/2$, where  $\sigma_I$ is the energy per
unit area of the domain wall $(I)$ defined in \sigmaI\ .
On the Minkowski side $a = 0$.
For the type $II$ domain wall
when the potential has $Z_2$ symmetry,
the energy per unit area
\sigmaII\ is $\sigma_{II}= 4\kappa^{-1}
|{\Lambda_{\pm\infty}/
 3}|^{1/2}$
and the fiducial observer is repelled on both sides of the
domain wall with the same acceleration  $a_{\pm \infty}\to
\kappa \sigma_{II}/4$ which resembles
remarkably the form for the acceleration for the
domain walls discussed in Ref.\refmark{\IPSK}.
In our case the domain wall also respects the
$Z_2$ symmetry, however, it is completely static
and its repulsive nature is due to
the AdS nature of the asymptotic space-time.

\REF{\HE}{ S. W. Hawking and G. F. R. Ellis, \sl The Large Scale Structure
of Space-Time \rm , Cambridge, 1973.}

\REF{\AIS}{S. J. Avis, C. J. Isham, and D. Storey, Phys. Rev. \bf D18 \rm
(1978) 3565.}

\REF{\DF}{P. C. W. Davies and S. A. Fulling, Proc. R. Soc. Lond., \bf 356 \rm ,
237  (1977).}

\chap{Anti-de~Sitter--Minkowski walls}

In many ways, the walls separating flat Minkowski space-time from AdS
are the most interesting.  For example, it is known that AdS has the
topology $S^{1}(time) \times \Re^{3}(space)$ and thus has closed
time-like curves. The common remedy for such loss of causality is
to unwrap the time direction and work on the covering space CAdS.
Nevertheless, this  does not allow AdS to be globally
hyperbolic; i.e. it has no Cauchy hypersurface and thus
boundary conditions must be imposed at spatial infinity
in order to properly specify the Cauchy problem
\refmark{\HE\ , \AIS}.
In the juxtaposition of AdS and Minkowski
space-times, one must consider
a proper formulation of the Cauchy problem in order
to quantize a field living on this manifold.
The problems of AdS in some ways are softened by the
presence of the Minkowski side, yet one unfortunately
cannot erase the problems associated with a
lack of a Cauchy surface.

As an indication of the interesting causal structure obtained
through the juxtaposition of AdS and Minkowski space-times,
consider the $1+1$ Minkowski-$AdS_{2}$
wall and place
an observer on the Minkowski side.
Now allow the observer to
send a moving mirror through the wall on
a geodesic. The mirror will
travel on a hyperbolic
trajectory as it falls
into the AdS space-time.
If the Minkowski observer sends massless radiation
at the mirror, s/he will receive more reflected  radiation out than
sent in due to the coupling of the mirror to the curved space-time and
the resulting particle creation\refmark{\BD}.
In this way the Minkowski observer could deduce the structure of the
space-time on the other side of the wall.
In addition, it is known that the energy radiated from the mirror
on a hyperbolic trajectory is zero\refmark{\DF} and thus the
stability of the wall is not compromised.

Note that the moving mirror will reach the end of the coordinates
$z,t$ within a finite proper time ({\it i.e.} these coordinates must be
extended in order to cover AdS).
However, the Minkowski observer
considers $t$ his/her proper time.
Such behaviour is true in the full $3+1$ case as well.
Therefore, to construct the causal structure of the
domain wall system, we must extend the coordinates
on the AdS side, which means
an extension onto a new half of AdS is necessary.  However, by allowing
for more AdS, we have added more effective gravitational mass to
the system which must be cancelled by another identical wall.
On the other side of the wall there is another
Minkowski space. Moving to the covering space to
avoid the closed time-like curves inherited from AdS
gives us an infinite tower of 2-wall systems.
The Penrose diagram for this configuration is shown in figure 6.
Of particular interest is the null separating the AdS patches.
It turns out that this surface corresponds to the Cauchy
horizon\refmark\CGS   with the metric closely related to the one at the
extremal  Reissner Nordst\"om  (RN)
black-hole horizon. Note the similarity of
the Penrose diagram for our domain wall and the one  of the
extremal (RN) black hole; the only major difference is
that in our case the singularity  of the RN black-hole inside
the Cauchy horizon is  replaced by an identical wall.
Space-time  induced by such domain walls thus
serves as an example of asymptotical Minkowski space-time with
Cauchy horizon but without any
singularity.
Further study of the    phenomena associated with the
Minkowski-AdS wall is underway\refmark{\CGS}.

\topinsert
Figure 6: Penrose diagram for the
covering space of the extended Minkowski-AdS domain wall
system.
The regions $M$ and $A$ are the Minkowski and AdS sides of
the wall.
The vertical line is the time-like line of the
domain wall.
The nulls separating AdS patches are
sights of instabilities.
The point $\Omega$ is on
one such null.  At this point, the observer experiences
the complete history of his/her preceeding Minkowski region.

\line{\hrulefill}
\endinsert

\chap{Summary}

We studied the field theoretic realization of a new
type of domain wall.  These walls separate
two maximally symmetric space-times of non-positive cosmological
constant where one or both is AdS and the other can be Minkowski.
These walls are found in $N=1,d=4$ supergravity and saturate
a positive energy/area theorem thus providing stability to the
classical configuration.  We classified three canonical systems
differing by the path in superpotential space traced out as the
scalar field interpolated from one vacuum to the other.
Equivalently,
the form of the conformal factor on the conformally flat metric
characterizes the three walls.
Examples were given illustrating the three walls as realized by
a particular superpotential.
And the motion of test particles living in the
background space-time induced by the walls was discussed.
 Finally, we
pointed out some interesting behaviour in regard to the
space-times induced from the
Minkowski-AdS walls.

We wish to thank R. Davis, S.-J. Rey, and H. H. Soleng for
enjoyable collaborations and many enlightening discussions.
M.C.'s work is supported in part by DOE and Texas
SSC funds.

\refout

\endpage

\bf Figure Captions \rm

Figure 1:  The path in superpotial space traversed as the
scalar field interpolates between degenerate vacua.
The wall is realized in both the global and local theories
for path (A) and just for the global theory in path (B).

Figure 2: The projection of a
typical scalar potential on a
particular complex $T$ direction
indicating the supersymmetric
minima ($D_{W}T = 0$) between which the domain wall
interpolates.  These minima are in general not degenerate.
However, since they are supersymmetric, when gravitational
energy is included they become energetically the same thus allowing for
domain wall solutions.

Figure 3:
Type $(I)$
conformal factor $A(z)$ for a space-time with $\Lambda_{-\infty}
= 0$
(Minkowski: $z<0$) separated by a domain wall from a space-time
with $\Lambda_{+\infty}
 < 0$ (AdS: $z>0$).  The wall, \ie\
 the
region over which the matter field $T$ changes is centered
at $z=0$  and has thickness
$\approx 200$ in $ \sqrt\kappa$ units.
The superpotential \superpotential\ has parameters $a^{2}=0,
b^{2}=0.1$ and $T$
interpolates between  $T_{-\infty}= 0 = a$
and $T_{\infty}= .318 \approx b$.

Figure 4:  Type $(II)$
Conformal factor $A(z)$ for a space-time with negative
cosmological constant separated by a domain wall from its mirror image
(\ie\ a $Z_{2}$ configuration).  The wall is centered at $z=0$ and has
thickness $\approx 200$ in $\sqrt\kappa$ units.
The superpotential
\superpotential\  has parameters $a^{2}=.025, b^{2}=0.1$.
and $T$
interpolates between  $T_{\mp\infty}= \pm .1598 \approx \pm a$.

Figure 5: Type $(III)$
conformal factor $A(z)$ for a space with negative
cosmological constant
separated by a domain wall from a space with a different
negative cosmological constant.  The superpotential $W$ never passes
through a zero as $T$ interpolates from one vacuum to another.
The domain wall is centered at $z=0$ and
has thickness $\approx 200$
where   $z$ is measured in $\sqrt\kappa$ units.
The singularity is at $z^{*}
\approx -5600$.
The superpotential
\superpotential\ has parameters $a^{2}=.025, b^{2}=0.1$
and $T$
interpolates between  $T_{-\infty}= .315 \approx b $
and $T_{\infty}= .160 \approx a$.

Figure 6: Penrose diagram for the
covering space of the extended Minkowski-AdS domain wall
system.
The regions $M$ and $A$ are the Minkowski and AdS sides of
the wall.
The vertical line is the time-like line of the
domain wall.
The nulls separating AdS patches are
sights of instabilities.
The point $\Omega$ is on
one such null.  At this point, the observer experiences
the complete history of his/her preceeding Minkowski region.

\end